\documentclass[runningheads]{llncs}
\usepackage[T1]{fontenc}
\usepackage{graphicx}
\usepackage{graphics}
\usepackage{cite}
\usepackage{amsmath,amssymb,amsfonts}
\usepackage{algorithm}
\usepackage{algorithmic}
\usepackage{graphicx}
\usepackage{textcomp}
\usepackage{array}
\usepackage{tabularx,booktabs}
\begin{document}
%
\title{An Unsupervised Framework for Joint MRI Super Resolution and Gibbs Artifact Removal}
\titlerunning{An Unsupervised Framework for Joint MRI SR and Gibbs Artifact Removal}
%
\author{Yikang Liu
\and Eric Z. Chen\index{Chen,Eric Z.} 
\and Xiao Chen
\and Terrence Chen
\and Shanhui Sun
}
\authorrunning{Y. Liu et al.}
%
\institute{United Imaging Intelligence, Cambridge, MA, USA\\
\email{\{yikang.liu, zhang.chen, xiao.chen01, terrence.chen, shanhui.sun\}@uii-ai.com}}
%
\maketitle              
\begin{abstract}
The k-space data generated from magnetic resonance imaging (MRI) is only a finite sampling of underlying signals. Therefore, MRI images often suffer from low spatial resolution and Gibbs ringing artifacts. Previous studies tackled these two problems separately, where super resolution methods tend to enhance Gibbs artifacts, whereas Gibbs ringing removal methods tend to blur the images. It is also a challenge that high resolution ground truth is hard to obtain in clinical MRI. In this paper, we propose an unsupervised learning framework for both MRI super resolution and Gibbs artifacts removal without using high resolution ground truth. Furthermore, we propose regularization methods to improve the model's generalizability across out-of-distribution MRI images. We evaluated our proposed methods with other state-of-the-art methods on eight MRI datasets with various contrasts and anatomical structures. Our method not only achieves the best SR performance but also significantly reduces the Gibbs artifacts. Our method also demonstrates good generalizability across different datasets, which is beneficial to clinical applications where training data are usually scarce and biased.

\end{abstract}
\section{Introduction}
\begin{figure}[!t]
    \centering
    \includegraphics[width=\linewidth,trim={0 80 300 0},clip]{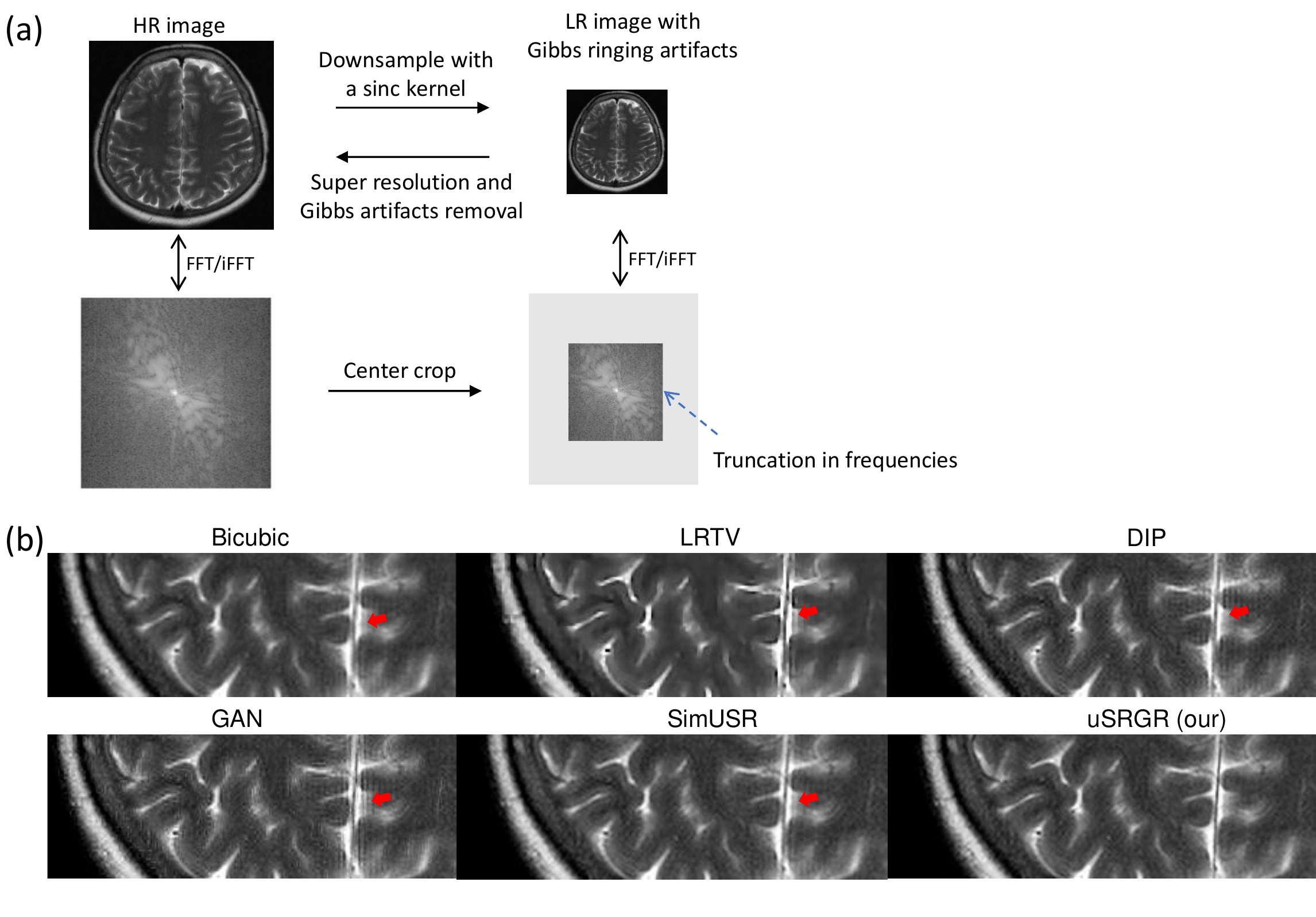}
    \caption{(a) An illustration of LR MRI data generation process. The acquired LR MRI image can be viewed as a result of a center crop of the underlying HR k-space, which leads to the common Gibbs ringing artifacts. Here HR image is used for demonstration purposes and is rarely available in practice. (b) Examples of SR (2x) and Gibbs artifacts (red arrows). No ground truth of HR image is available.}
    \label{fig:demo}
\end{figure}

The super resolution (SR) for magnetic resonance imaging (MRI) images is different from natural images due to the distinct image generation process. Due to various constraints such as hardware limitations and acquisition time, the k-space data generated from MRI is only a finite sampling of spatial frequencies, which leads to the loss of high spatial frequencies. The generated low resolution (LR) images make subtle structures and boundaries hardly distinguishable. Another common MRI image artifact called Gibbs ringing often arises in practice due to abrupt cutoff in spatial frequency (Fig.~\ref{fig:demo}a), which may appear as anatomical structures and cause misinterpretation. Therefore, SR for MRI images is more complex than natural images since solving these two problems individually often leads to a dilemma. SR tends to enhance Gibbs ringing artifacts (Fig.~\ref{fig:demo}b) while Gibbs ringing artifact removal (deGibbs) algorithms might generate blurry results (Fig.~\ref{fig:ablation}d). Furthermore, the artifact removal algorithms are often performed on the original resolution \cite{Kellner2016,Zhang2019c, Muckley2019}. Since both problems are caused by finite k-space signals, we propose to solve them simultaneously with a unified deep learning framework. However, there are several challenges. First, supervised learning becomes a nontrivial task due to the lack of high resolution (HR) ground-truth MRI images. Second, there is usually a distribution mismatch between training and testing data, since it is impractical to collect training images of all contrasts, anatomies, diseases, etc. Third, it is crucial for clinical diagnosis that the model should not introduce additional artifacts into SR images.

To address these problems, we propose an \textbf{U}nsupervised learning framework for joint MRI \textbf{S}uper \textbf{R}esolution and \textbf{G}ibbs artifact \textbf{R}emoval (uSRGR), which utilizes specifically designed loss constraints for image fidelity and model generalizability. The model learns the inverse process of downsampling MRI images, which is to crop out the peripheral (high-frequency) part of the frequency space (obtained by applying Fourier transform to MRI images) with a box window and leads to reduced image resolution and Gibbs artifacts (Fig.~\ref{fig:demo}a). 

To avoid using HR images, LR images $\boldsymbol{I}_{\text{LR}}$ are downsampled into images of lower resolution $\boldsymbol{I}_{\text{LR}'}$, and the model is trained to fit $\boldsymbol{I}_{\text{LR}'}$ to $\boldsymbol{I}_{\text{LR}}$. At the same time, the model predicts HR images $\Hat{\boldsymbol{I}_{\text{HR}}}$ from $\boldsymbol{I}_{\text{LR}}$ images. To ensure the fidelity of prediction, the difference between downsampled $\Hat{\boldsymbol{I}_{\text{HR}}}$ and $\boldsymbol{I}_\text{LR}$ is minimized. Moreover, 
since cropping peripheral parts of frequency domain (hereinafter referred to as `f-cropping') has a more global effect than the downsampling process in natural images (e.g. convolution with a kernel of a finite size), we utilize large 1D convolutional kernels to model such an effect. Furthermore, to improve the model generalizability, we regularize the model to match a Sinc convolution process that downsamples $\boldsymbol{I}_{\text{HR}}$ to $\boldsymbol{I}_{\text{LR}}$, since Sinc convolution in the spatial domain is equivalent to f-cropping. In summary, our contributions are as follows:
\begin{itemize}
\item We propose a novel unsupervised deep learning framework for joint SR and Gibbs artifact removal without using HR MRI images.
\item We demonstrate an innovative training paradigm that regularizes the model to a Sinc deconvolution process to improve learning generalization. 
\end{itemize}

\section{Related Works}
Many supervised deep learning methods have been proposed for image SR \cite{Mahapatra2019, Pham2019, Chen2018, Umehara2018, Lyu2020, Huang2020, Ravi2019}. Some unsupervised deep learning methods have also been proposed \cite{Ravi2019, Menon2020, 8575264, kim2020unsupervised}, where paired LR and HR images are not required for training but HR images are still needed to learn the HR image manifold. 
Without using HR images, several unsupervised or self-supervised SR methods have been proposed for natural images. Deep image prior (DIP) \cite{Lempitsky2018, Mataev2019} uses a network structure as a prior to constrain HR prediction. Zero-shot SR (ZSSR) \cite{Shocher2017} is another online-learning method that is trained in a self-supervised manner using a pair of the downsampled and original LR images. Following ZSSR, SimUSR \cite{Ahn2020} is trained offline on many image pairs of downsampled and original LR images to reduce inference time. GAN based methods have also been proposed for unsupervised SR \cite{Chen_2020_CVPR_Workshops,Ravi2019, Menon2020, 8575264, kim2020unsupervised}. However, GAN tends to generate new artifacts \cite{Zhang2019}. In the absence of a supervised loss from HR images, there is a risk that the GAN model may generate artifacts that result in clinical misdiagnosis.  

As for Gibbs removal, traditional methods aim to minimize oscillation in images with image filtering \cite{Gottlieb1997, Jerri2000LanczosLikeF}, Gegenbauer polynomial reconstruction \cite{1000255}, frequency space extrapolation regularized by total variation \cite{Block2008}, and sub-voxel shift resampling \cite{Kellner2016}. Since image details can also be smoothed as oscillation, these methods blur images to varying degrees \cite{Kellner2016}. Recently proposed deep learning models \cite{Zhang2019c, Muckley2019} are trained on pairs of artifactual and clean images, where artifactual images were synthesized by f-cropping HR images \cite{Muckley2019,Zhang2019c}. Similar to our method, \cite{Zhang2019c} also tried to model the inverse of the f-cropping process. However, it is different from our work in multiple ways: first, it is a supervised learning method; second, distribution mismatch between training and test data is not considered; third, \cite{Zhang2019c} crops out $50\%$ of the original frequency space as the downsampling operation whereas we crop out $75\%$. 

\section{Methods}
\subsection{Problem Formulation}
A LR image $\boldsymbol{I}_{LR}\in\mathbb{R}^{n\times m}$ can be generated by downsampling a HR image $\boldsymbol{I}_{HR}\in\mathbb{R}^{N\times M}$ as $\boldsymbol{I}_{LR} = d(\boldsymbol{I}_{HR}) + \epsilon$,
where $d$ is a downsampling operation, and $\epsilon$ is the noise. For MRI, $d$ is f-cropping $d_{fc}$, and $\boldsymbol{I}_{LR}$ often bears Gibbs ringing artifacts. Therefore, the joint task of SR and deGibbs can be formulated as an inverse problem where we need to approximate the inverse operation of the downsampling operation $d_{fc}$ with a deep neural network $f_{\theta}$: 
$\boldsymbol{I}_{HR} = f_{\theta}(\boldsymbol{I}_{LR})$.

\subsection{An Unsupervised Learning Framework}
\begin{figure}[!tbp]
    \centering{
    \includegraphics[width=\linewidth,trim={0 5 0 0},clip]{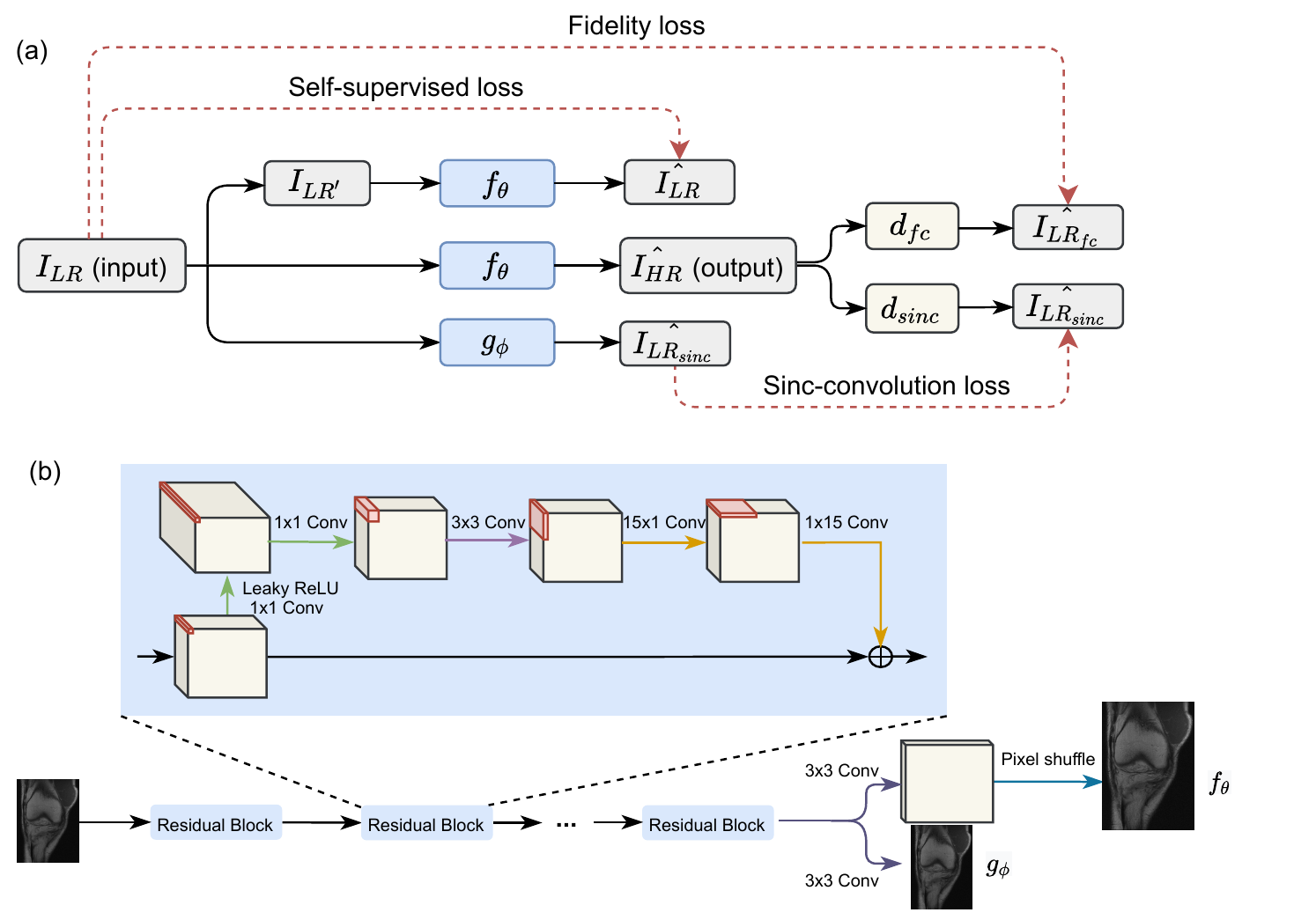}
    }
    \caption{(a) The proposed unsupervised framework uSRGR for MRI SR and Gibbs removal. (b) The network architecture used in uSRGR. The networks $f_{\theta}$ and $g_{\phi}$ share the same architecture except for the last pixel shuffle layer due to different output sizes.}
    \label{fig:framework}
\end{figure}

Without HR ground truth, we propose to regularize SR neural network $f_{\theta}$ with three constraints given only LR images (Fig.~\ref{fig:framework}a): a fidelity constraint, a self-supervision constraint, and a Sinc-convolution constraint.

{\bf Self-Supervision Constraint: }
Based on the assumption that the inverse operation of downsampling $d_{fc}$ is locally consistent across multiple scales \cite{Ahn2020}, we downsampled the LR image $\boldsymbol{I}_{LR}$ with $d_{fc}$ to a lower resolution $\boldsymbol{I}_{LR'}$ and define the self-supervision loss $\mathcal{L}_{ss}$ on pairs of $\boldsymbol{I}_{LR'}$ and $\boldsymbol{I}_{LR}$:
\begin{equation}\label{ss}
    \mathcal{L}_{ss}(\theta) = \sum_{i} \mathcal{L}_{d}(f_{\theta}(\boldsymbol{I}_{LR'}^{(i)}), \boldsymbol{I}_{LR}^{(i)})
\end{equation}

{\bf Fidelity Constraint: }
Signal fidelity in medical images is crucial in clinical settings, especially for parametric mappings where the intensities bear physical and physiological meanings. We downsample the predicted HR image $\Hat{\boldsymbol{I}_{HR}}$ with $d_{fc}$ and minimize its difference with the original LR image $\boldsymbol{I}_{LR}$ as
\begin{equation}\label{fid-loss}
    \mathcal{L}_{f}(\theta) = \sum_{i} \mathcal{L}_{d}(d_{fc}(f_{\theta}(\boldsymbol{I}_{LR}^{(i)})), \boldsymbol{I}_{LR}^{(i)}).
\end{equation}

{\bf Sinc-Convolution Constraint: }
We leverage the knowledge about the generation of LR MRI images and introduce another regularization on the f-cropping process $d_{fc}$. It enforces the model $f_{\theta}$ to learn the inverse of the underlying downsampling process $d_{fc}$ and prevents model overfitting on spurious image features. Thus the model can generalize better on out-of-distribution data. 

The central cropping in frequency domain is mathematically equivalent to convolution with a Sinc kernel of an infinite size in the spatial domain. Therefore, if we downsample $f_{\theta}(\boldsymbol{I}_{LR})$ by convolution with a Sinc kernel of a \emph{finite} size, it should be close to $\boldsymbol{I}_{LR}$, which provides another supervision on the mapping from $\boldsymbol{I}_{LR}$ to $\boldsymbol{I}_{HR}$. 
To compensate the difference between $d_{fc}$ and $d_{sinc}$ caused by the finite Sinc kernel, we first pretrain another network $g_{\phi}$ that maps images downsampled with f-cropping $d_{fc}(\boldsymbol{I}_{LR})$ to the ones downsampled with Sinc convolution $d_{sinc}(\boldsymbol{I}_{LR}))$ on a dataset of LR images $\boldsymbol{I}_{LR}$ with the loss function:
\begin{equation}\label{g}
    \mathcal{L}({\phi}) = \sum_{i} \mathcal{L}_{d}(g_{\phi}(d_{fc}(\boldsymbol{I}_{LR})), d_{sinc}(\boldsymbol{I}_{LR}))
\end{equation}
Then the Sinc-convolution constraint on the model $f_{\theta}$ is
\begin{equation}\label{sinc}
    \mathcal{L}_{sinc}(\theta) = \sum\max(\mathcal{L}_{d}(d_{sinc}(f_{\theta}(\boldsymbol{I})),g_{\phi}(\boldsymbol{I})), a),
\end{equation}
where $a=0.001$ is a small number to compensate the prediction error in $g_{\phi}$ and $\boldsymbol{I}$ is sampled from $\boldsymbol{I}_{LR}$ and $\boldsymbol{I}_{LR'}$ images.

Taken together, the total loss function for the SR model $f_{\theta}$ is
\begin{equation}
    \mathcal{L}(\theta) = \mathcal{L}_{ss} + \beta \mathcal{L}_f+ \gamma \mathcal{L}_{sinc}
\end{equation}
, 
where $\beta=1$ and $\gamma=0.5$ are weights. We use a combination of L1 loss and multi-scale SSIM loss \cite{Zhao2015} for $\mathcal{L}_{d}$. The model $g_{\phi}$ is also fined tuned with the loss function Eq.~\ref{g} during the training of the SR model $f_{\theta}$, where both $\boldsymbol{I}_{LR}$ and $\Hat{\boldsymbol{I}_{HR}}$ (i.e. $f_{\theta}(\boldsymbol{I}_{LR})$) are used. In this way, the models $f_{\theta}$ and $g_{\phi}$ need to come to agreements on both HR and LR scales, which provides extra regularization for $f_{\theta}$ in addition to Eq.~\ref{fid-loss}.

{\bf Network: }
For efficient computation, we adopt a simple residual network architecture with wide activation\cite{Yu2020} for both $f_{\theta}$ and $g_{\phi}$ (Fig.~\ref{fig:framework}b). To model the global effects of Gibbs artifacts, we utilized two 1D large-kernel convolutions to approximate a 2D large-kernel convolution. We employ a pixel shuffle layer in $f_{\theta}$ that takes LR images as inputs, which is more efficient than taking interpolated LR images as inputs \cite{Shi2016}.

\section{Experiment setup}
{\bf Datasets}
We included four datasets (T1, T1pre, T1post, T2) from the public (under a non-exclusive, royalty-free license) fastMRI brain images \cite{DBLP:journals/corr/abs-1811-08839} and two datasets (PDFS and PD) from fastMRI knee images \cite{DBLP:journals/corr/abs-1811-08839}. The fastMRI dataset\footnote{https://fastmri.org/dataset/} \cite{DBLP:journals/corr/abs-1811-08839} contains MRI k-space data of brain and knee images. We reconstructed images from k-space data and cropped them to 320 x 320. The brain images have four contrasts: T1 (denoted as \textit{brain-T1}), T1 before and after injection of contrast agent (denoted as \textit{brain-T1pre} and \textit{brain-T1post}, respectively), T2 (denoted as \textit{brain-T2}), and T2 FLAIR. T2-FLAIR images were not used due to high motion artifacts. The knee images have two contrasts: proton-density weighting with and without fat suppression, denoted as \textit{knee-PDFS} and \textit{knee-PD}, respectively. The imaging parameters can be found in \cite{DBLP:journals/corr/abs-1811-08839}. brain-T2, brain-T1post, brain-T1pre, brain-T1, knee-PD, knee-PDFS contains 21500, 123600, 32700, 9200, 5000, and 9900 images for testing. 267800 brain-T2 and 48400 knee-PD images were used for training separately, with 60000 brain-T2 and 5000 knee-PD images used respectively for validation. 

We also included two private cardiac cine MRI datasets (details can be found in supplementary material), where retroCine dataset contains 2,955 images from 51 subjects and rtCine dataset contains 57 images from two subjects. retroCine data of 51 subjects (total 2955 images) were acquired on a 3T scanner (uMR 790 United Imaging Healthcare, Shanghai, China) and phased-arrayed coils using a bSSFP sequence. rtCine data of 57 images were acquired from two patients using the same equipment and a bSSFP sequence using variable Latin Hypercube undersampling \cite{Lyu_2019_ISMRM}. Imaging parameters were: matrix size = 192 x 180, TR = 2.8 $ms$, TE = 1.3 $ms$, spatial resolution = 1.82 x 1.82 $mm^{2}$, and temporal resolution = 34 $ms$ and 42 $ms$ for retroCine and rtCine. Acquisition of both retroCine and rtCine was approved by a local institutional review board.

{\bf Compared methods}
We compared uSRGR with four state-of-the-art SR methods. LRTV \cite{Shi2015} is a MRI SR method using low-rank and total variation regularization\footnote{https://bitbucket.org/fengshi421/superresolutiontoolkit}. Three unsupervised deep learning methods are included: DIP \cite{Lempitsky2018}, SimUSR \cite{Ahn2020}, and a GAN-based method extended from \cite{10.5555/2969033.2969125,Bell-Kligler2019,isola2018imagetoimage}. In the GAN-based method, we used a patch discriminator to distinguish patches derived from LR inputs and HR predictions and the objective function is $f = \arg\min_{f} \{(\max_{D} \mathbb{E}_{x \sim patches(I_{LR})}|D(x)-1| + |D(f(x))|) + \lambda\mathcal{L}_{d}(ds(f(I_{LR})), I_{LR})\}$, where $f$ is a network with the same architecture as in Fig~\ref{fig:framework}. 
\begin{table}[ht]

    \centering
\begin{tabular}{ll}
    \resizebox{0.35\columnwidth}{!}{%
    \begin{tabular}{cc}
         Layer Name & Parameter Dimension \\
         \hline
         conv2d & $6N \times N \times 1 \times 1$\\
         LeakyReLU & \\
         conv2d & $\text{int}(4.8N) \times 6N \times 1 \times 1$\\
         conv2d & $\text{int}(4.8N) \times \text{int}(4.8N) \times 3 \times 3$\\
         conv1d & $\text{int}(4.8N) \times \text{int}(4.8N) \times 15 $\\
         conv1d & $N \times \text{int}(4.8N) \times 15 $\\
    \end{tabular}
    }
&
    \resizebox{0.32\columnwidth}{!}{%
    \begin{tabular}{c c}
         Layer Name & Parameter Dimension \\
         \hline
         conv2d & $6N \times N \times 1 \times 1$\\
         LeakyReLU & \\
         conv2d & $\text{int}(4.8N) \times 6N \times 1 \times 1$\\
         conv2d & $N \times \text{int}(4.8N) \times 3 \times 3$\\
    \end{tabular}
    }
\end{tabular}
\caption{Residual blocks with (left) and without (right) 1D convolutions.}
\label{tab:resblock}
\end{table}

{\bf Implementation Details} The uSRGR model included 12 residual blocks ($N=80$) with 1D convolutional layers (Table~\ref{tab:resblock} and Fig. 2) and two $3 \times 3$ 2D convolutional layers before and after the residual blocks. For uSRGR-1Dconv and other baselines, we removed the 1D convolutions in the residual blocks (Table~\ref{tab:resblock}) and increased the number of channels $N$ to 135, so that they had comparable number of parameters (uSRGR: 4.2M parameters; others: 4.4M parameters).

All models were implemented with PyTorch 1.5 and trained on NVIDIA V100. We used Adam optimizer \cite{DBLP:journals/corr/KingmaB14} with a learning rate of $10^{-4}$. Batch size was 4 and images were cropped into $160 \times 160$ patches. All models were trained for 200 epochs. Training losses converged well and no overfitting was observed.

\section{Results and Conclusion}
\begin{figure}[!tbp]
    \centering
    \includegraphics[width=0.93\linewidth,trim={0 10 0 0},clip]{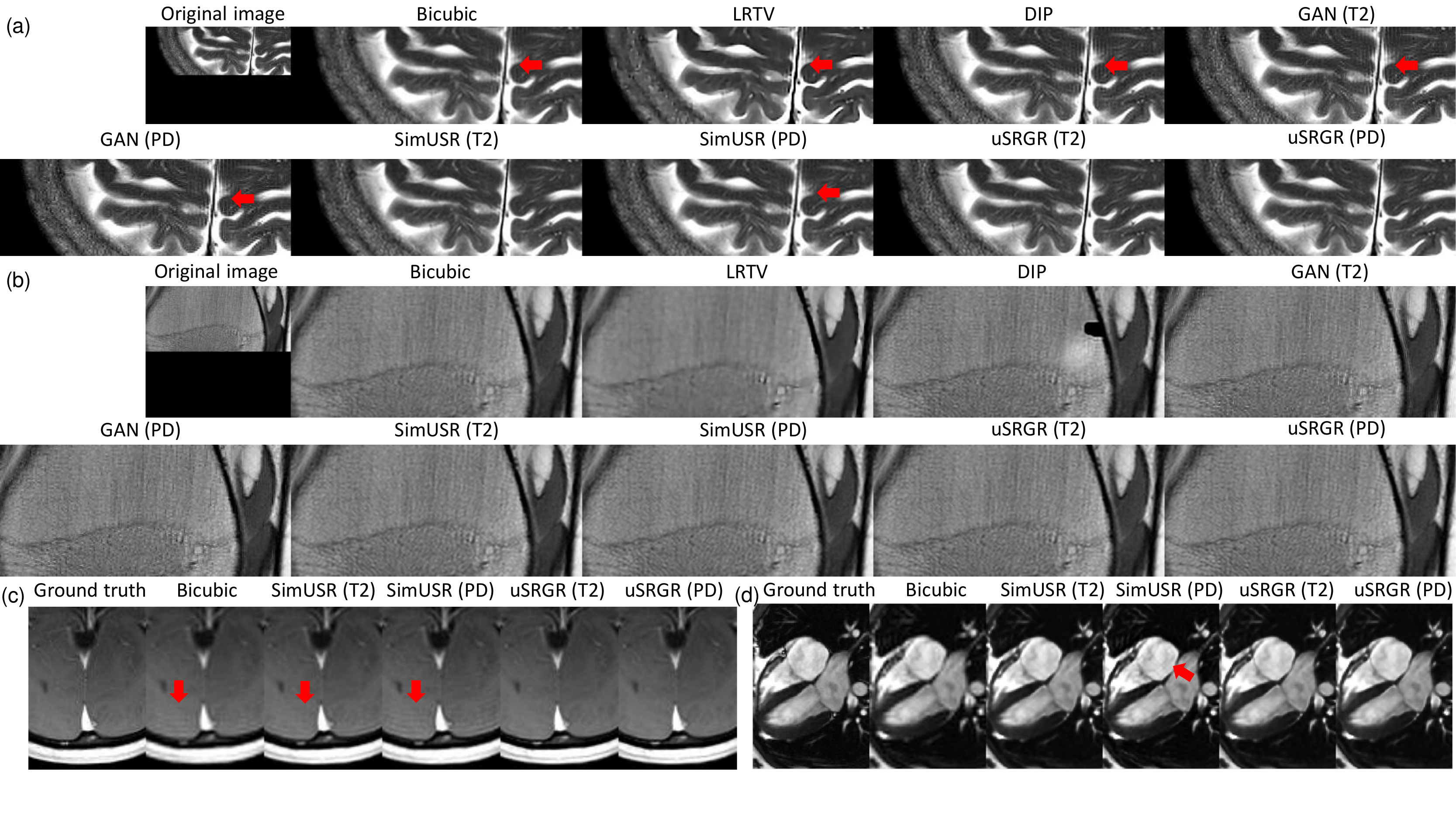}
    \caption{Comparison of different methods tested on (a), (b) the original datasets (brain-T2 and knee-PD) and (c), (d) synthetic LR datasets (brain-T1post and retroCine). T2 and PD in figure titles indicate training datasets. Red arrows indicate Gibbs artifacts.}
    \label{fig:baselines}
\end{figure}
{\bf Quantitative comparisons: } 
Since HR MRI images are not available, we used the original resolution images as the evaluation reference for quantitative comparisons. We synthesized LR images (2x downsampling) from original images using f-cropping to mimic the MRI downsampling process. Furthermore, to test the generalizability of different methods, we trained models on brain-T2 and knee-PD datasets respectively and tested on all eight datasets, where brain-T2 and knee-PD were split into training, validation, and testing sets. The test data with the same imaging sequence and organ as the training data is referred to as the \textit{within-distribution}(WD) data and otherwise as the \textit{out-of-distribution}(OD) data.
Table~\ref{tab:benchmark} shows quantitative results for different SR methods. Our method performed significantly better than other methods on both WD and OD data ($p<0.05$). No significant difference ($p>0.2$) between uSRGR models trained on brain-T2 and knee-PD was found on any test dataset, indicating its good generalizability.  

\begin{table}[!tbp]
    \caption{Evaluation on synthetic LR images with original images as ground truth. Rows indicate the methods (training datasets) and columns indicate test datasets. } 
    \label{tab:benchmark}
    \centering
  \begin{tabular}{lcccc}

    \toprule
    PSNR/SSIM & brain-T2 & brain-T1 & brain-T1post & brain-T1pre \\
    \hline
    Bicubic &  32.08/0.912  & 35.01/0.940  & 34.94/0.942 &  34.85/0.933 \\
    LRTV & 29.64/0.898 & 32.88/0.931 & 32.65/0.934 & 32.67/0.925 \\
    \hline
    DIP & 34.52/0.919 & 37.93/0.943  & 37.82/0.946 & 37.72/0.929 \\
    GAN (T2) & 33.56/0.905 & 36.75/0.931 & 36.39/0.934 & 36.16/0.925 \\
    GAN (PD) & 33.24/0.905 & 36.73/0.932 & 36.25/0.936 & 36.15/0.928 \\
    simUSR (T2) & 36.39/0.938 & 39.67/0.952 & 39.12/0.956 & 38.60/0.948 \\
    simUSR (PD) & 35.80/0.936 & 39.91/0.957 & 38.92/0.960 & 38.38/0.951 \\
    \hline
    uSRGR (T2) & \textbf{36.87}/\textbf{0.949} & \textbf{40.48/0.963} & \textbf{39.65/0.962} & \textbf{38.98/0.959} \\
    uSRGR (PD) & \textbf{36.86/0.945} & \textbf{40.48}/\textbf{0.964} & \textbf{39.63}/\textbf{0.960} & \textbf{38.97}/\textbf{0.960} \\
    \hline
    uSRGR-fid (T2) &36.43/0.939 & 40.05/0.951 & 39.17/0.955 & 38.42/0.951\\ 
    uSRGR-fid (PD) &36.40/0.935 & 39.99/0.956 & 39.25/0.959 & 38.60/0.954\\
    uSRGR-sinc (T2) &36.46/0.941 & 40.25/0.953 & 39.43/0.959 & 38.54/0.953\\
    uSRGR-sinc (PD) &35.68/0.937 & 40.06/0.957 & 38.91/0.960 & 38.39/0.952\\
    uSRGR-1Dconv (T2) &36.73/0.942 & 40.27/0.955 & 39.538/0.960 & 38.813/0.955 \\
    uSRGR-1Dconv (PD) &36.70/0.923 & 40.31/0.959 & 39.63/0.960 & 38.97/0.958\\
    deG+SR (T2) & 36.12/0.910 & 39.02/0.941 & 38.64/0.952 & 38.20/0.944\\
    deG+SR (PD) & 36.08/0.913 & 38.87/0.936 & 38.55/0.953 & 38.13/0.941\\
    \toprule
    \toprule
    PSNR/SSIM &knee-PD & knee-PDFS & retroCine & rtCine \\
    \hline
    Bicubic & 34.07/0.888 & 34.32/0.823 & 30.97/0.896 & 31.49/0.914 \\
    LRTV & 32.28/0.874 & 33.09/0.807 & 27.97/0.877 & 28.69/0.895 \\
        \hline
    DIP & 34.81/0.897 & 34.14/0.834 & 33.91/0.913 & 34.27/0.932 \\
    GAN (T2) & 33.90/0.884 & 33.84/0.815 & 32.27/0.896 & 32.87/0.913 \\
    GAN (PD) & 33.78/0.883 & 33.88/0.815 & 31.97/0.896 & 32.58/0.913 \\
    simUSR (T2) & 36.98/0.916 & 36.13/0.854 & 35.23/0.944 & 36.06/0.954 \\
    simUSR (PD) & 37.29/0.922 & 36.15/0.856 & 35.47/0.948 & 36.42/0.960 \\
    \hline
	uSRGR (T2) & \textbf{37.59/0.930} & \textbf{36.50/0.863} & \textbf{35.81/0.953} & \textbf{36.74/0.965} \\
    uSRGR (PD) & \textbf{37.61}/\textbf{0.932} & \textbf{36.52}/\textbf{0.863} & \textbf{35.81}/\textbf{0.954} & \textbf{36.75}/\textbf{0.966} \\
    \hline
    uSRGR-fid (T2) & 37.09/0.919 & 36.14/0.852 & 35.32/0.945 & 36.16/0.956\\
    uSRGR-fid (PD) & 37.00/0.918 & 36.05/0.854 & 35.30/0.946 & 36.23/0.957\\
    uSRGR-sinc (T2) & 37.24/0.922 & 36.16/0.852 & 35.57/0.949 & 36.37/0.959\\
    uSRGR-sinc (PD) & 37.28/0.923 & 36.26/0.856 & 35.54/0.947 & 36.41/0.960\\
    uSRGR-1Dconv (T2) &37.35/0.923 & 36.33/0.856 & 35.62/0.949 & 36.55/0.961\\
    uSRGR-1Dconv (PD) & 37.41/0.923 & 36.33/0.858 & 35.65/0.950 & 36.56/0.961\\
    deG+SR (T2) &36.54/0.905 & 35.93/0.850 & 34.97/0.932 & 35.78/0.939\\
    deG+SR (PD) &36.69/0.910 & 35.88/0.851 & 35.12/0.935 & 35.91/0.938 \\
    \bottomrule
  \end{tabular}
\end{table}

{\bf Qualitative comparisons:} 
For qualitative evaluation, besides the synthetic LR images, we also trained all models on the original images to mimic the real MRI SR setting, where no HR ground truth is available. Similarly, all models were trained on brain-T2 and knee-PD datasets respectively and tested on all eight datasets to evaluate their generalizability.
Fig.~\ref{fig:baselines} shows that LRTV produces over-smoothed images while DIP generates hole-like artifacts and GAN also predicts dot and stripe artifacts. Moreover, these methods all enhance Gibbs artifacts. These observations are consistent with low PSNR/SSIM in Table~\ref{tab:benchmark}. simUSR removes Gibbs artifacts on WD data (simUSR(T2) on brain-T2) but fails to generalize on OD data (simUSR(PD) on brain-T2), also indicated by the significant difference ($p<0.05$) in PSNR/SSIM between simUSR(T2) and simUSR(PD) tested on WD and OD data in Table~\ref{tab:benchmark}. simUSR(PD) also failed to remove Gibbs artifacts in brain-T1post and retroCine images. In contrast, uSRGR generated sharper images with significantly ($p<0.05$) reduced Gibbs artifacts consistently across WD and OD data. 

{\bf Ablation Studies:} 
\begin{figure}[!tbp]
    \centering
    \includegraphics[width=\linewidth,trim={0 280 240 0},clip]{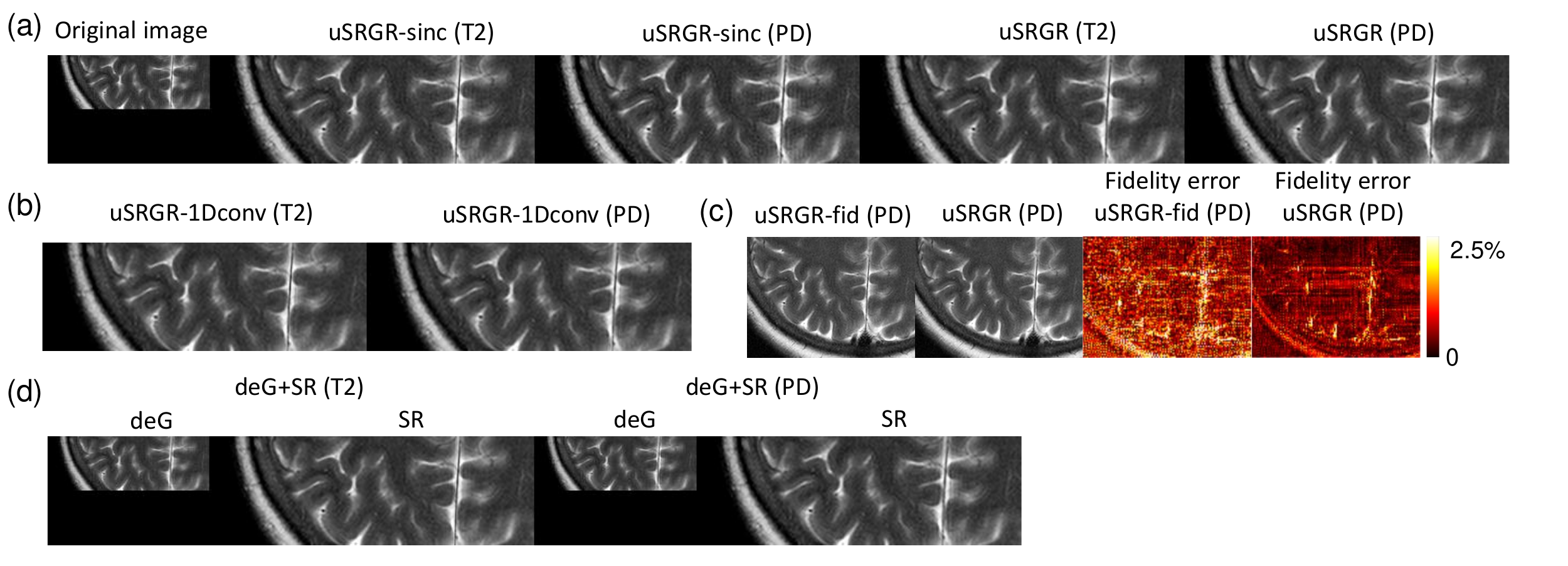}
    \caption{Ablation studies tested on brain-T2. 
    }
    \label{fig:ablation}
\end{figure}
We removed individual components in uSRGR: the fidelity constraint, the Sinc-convolution constraint, and the 1D convolutional layers, denoted as uSRGR-fid, uSRGR-sinc and uSRGR-1Dconv, respectively. As shown in Table~\ref{tab:benchmark}, removing those components individually leads to significantly lower PSNR/SSIM across all datasets ($p<0.05$). 
Removing the Sinc-convolution constraint resulted in more Gibbs artifacts (Fig.~\ref{fig:ablation}a), especially when trained on knee-PD and tested on brain-T2. This is consistent with our motivation that regularization based on the f-cropping process helps the model to learn the inverse process and thus improves the model generalizability. This also boosts the performance on WD data, since the models in Eq.\ref{ss} and Eq.\ref{fid-loss} operate on different image scales. 
Removing the fid loss significantly increases the fidelity error between the downsampled HR and the original image as in Fig.~\ref{fig:ablation}c (PSNR on all datasets: 49.2 vs. 44.7 with and without fid loss, $p<0.001$).  
Removing the 1D convolutions resulted in more blurry SR images (Fig.~\ref{fig:ablation}b). 

To demonstrate the benefit of joint SR and deGibbs, we trained separate models for deGibbs and SR (denoted as deG+SR). The same network $g_{\phi}$ was trained for deGibbs on image pairs downsampled with $d_{cubic}$ and $d_{fc}$. Then the separately trained network $f_{\theta}$ was applied for SR. Table~\ref{tab:benchmark} shows that deG+SR has lower PSNR/SSIM than uSRGR on all datasets ($p<0.05$). Fig.~\ref{fig:ablation}d shows that the separate deGibbs model blurred LR images and resulted in blurry SR images, indicates the advantage of combining deGibbs and SR in one framework.

\section{Limitations and Future Work}
This paper's aim is to develop an unsupervised deep learning framework for joint MRI SR and Gibbs artifact removal with good generalizability. Therefore, we did not put much effort into testing various loss functions that measure difference between two images (e.g. perceptual loss~\cite{johnson2016perceptual}), which may improve quality of predicted SR images. In addition, since Gibbs artifact has heavy effects on certain applications such as fiber tracking in diffusion tensor imaging, it would be beneficial to test the proposed method on such applications.

\section{Summary}
We proposed an unsupervised deep learning framework for MRI SR and Gibbs artifacts removal without using HR images as ground truth. The proposed method shows superior performance and generalizability than other unsupervised SR methods across various MRI datasets.

%
%
%
\bibliographystyle{splncs04}
\bibliography{ref}
\end{document}